\documentclass[%
 reprint,
superscriptaddress,
 amsmath,amssymb,
 aps, prl
]{revtex4-2}

\usepackage{graphicx}
\usepackage{dcolumn}
\usepackage{bm}

\begin{document}

\preprint{APS/123-QED}

\title{Four phonon-dominated near-field radiation in weakly anharmonic polar materials}

\author{Dudong Feng}
\affiliation{%
School of Mechanical Engineering and the Birck Nanotechnology Center, Purdue University, West Lafayette, Indiana 47907-2088, USA
}%

\author{Xiaolong Yang}
\altaffiliation[D. Feng and X. Yang contributed equally to this work]{}
\altaffiliation[X. Yang current at]{ College of Physics, and Center of Quantum Materials and Devices, Chongqing University, Chongqing 401331, China}
\affiliation{%
School of Mechanical Engineering and the Birck Nanotechnology Center, Purdue University, West Lafayette, Indiana 47907-2088, USA
}%

\author{Zherui Han}
\affiliation{%
 School of Mechanical Engineering and the Birck Nanotechnology Center, Purdue University, West Lafayette, Indiana 47907-2088, USA
}%

\author{Xiulin Ruan}%
 \email{ruan@purdue.edu}
\affiliation{%
School of Mechanical Engineering and the Birck Nanotechnology Center, Purdue University, West Lafayette, Indiana 47907-2088, USA
}%


\date{\today}

\begin{abstract}
Inelastic scattering processes typically introduce friction among carriers and reduce the transport properties of photons, phonons, and electrons. However, we predict that in contrast to the role in reducing thermal conductivity, four-phonon scattering dominates near-field radiative heat transfer (NFRHT) in both boron arsenide~(BAs) and boron antimonide. Including four-phonon scattering results in a nearly 400-fold increase in the total heat flux between two BAs thin-films compared to three-phonon scattering alone. This non-intuitive enhancement arises from the large number of NFRHT channels activated by four-phonon scattering outcompete the effect of decreased coupling strength of surface phonon polaritons at the resonance frequency. Additionally, we point out that four-phonon scattering to decrease NFRHT in certain other systems.
\end{abstract}

\maketitle

Enabled by coupled surface phonon-polaritons (SPhPs) at nanoscales, near-field thermal radiation between polar dielectrics achieves orders of magnitudes enhancement beyond the blackbody limit governed by Planck's law~\cite{greffet2002NMTE,zhang2020nano,sheng2009nanolet}. Such super-Planckian thermal radiation has been demonstrated experimentally~\cite{sheng2009nanolet,park2018prl,zhang2020nano}, leading to the developments of energy harvesting~\cite{Feng2021, chen2003apl}, thermal management~\cite{polini2017prl} and modulation~\cite{feng2021apl, fan2010prl}, and near-field imaging and sensing~\cite{Caldwell2015np, Greffet2002nature, Hillenbrand2002nature}. 

As hybrid quasiparticles formed by the strong coupling of surface electromagnetic (EM) modes and localized optical phonons, the phonon nature of SPhPs has been little investigated in the scenario of near-field radiative heat transfer (NFRHT)~\cite{fan2010prl}. Revealed by the dispersion relation, SPhPs can only be excited within the Reststrahlen band, which is bounded by the frequency of the zone-center transverse optical (TO), $\omega_\mathrm{TO}$, and longitudinal optical (LO) phonon modes, $\omega_\mathrm{LO}$. The dielectric function at this range can be effectively described by the Lorentz oscillator model, and the damping factor is the phonon linewidth, which manifests the phonon anharmonicity of the infrared (IR) active phonon at the zone center~\cite{yang2020prb}. According to fluctuational electrodynamics, the coupling strength of SPhPs is characterized by the damping factor, which explicitly connect the near-field radiative heat flux with phonon scattering processes.

While the lowest-order theory of phonon-phonon scattering were historically sufficient for explaining material anharmonicity~\cite{Lindsay2013prl, Broido2007APL}, considering higher-order intrinsic scattering, particularly four-phonon scattering, has significantly improved agreement between theoretical predictions and experimental measurements recently. Boron Arsenide (BAs) was predicited to exhibit high thermal conductivity of 2,200~W/m~K due to weak three-phonon scattering~\cite{Lindsay2013prl}. Later, a general theory of four-phonon scattering was introduced~\cite{feng2016prb}, predicting a reduced thermal conductivity of 1,400~W/m~K for BAs at room temperature~\cite{feng2017prb}. This significant reduction is due to restricted three-phonon scattering space making four-phonon scattering relativiely large~(referred to as weakly anharmonic polar materials). Three subsequent experiments confirmed BAs thermal conductivities of 1,000-1,300~W/m~K~\cite{Ren2018science,Hu2018science,David2018science}. BAs has also drawn significant attention for its high bi-polar mobilities, which were first predicted~\cite{Liu2018PRB} and later experimentally confirmed~\cite{chen2022science,Yue2022science}. Besides thermal conductivity, four-phonon scattering was recently shown to be significant in IR~\cite{yang2020prb, Tong2020prb} and Raman spectra~\cite{Han2022prl} of many materials, by resolving previous experiment-theory discrepancies for these materials. While extensive studies have delved into the influence of four-phonon scattering on related properties of weakly anharmonic polar materials~\cite{yang2020prb, Broid2020PRX, Xia2020PRX}, its impact on NFRHT remains unexplored.

In this Letter, we predict that four-phonon scattering can have a dominant contribution to NFRHT over three-phonon scattering in two prototype weakly anharmonic polar materials: BAs and Boron Antimonide (BSb). By combining first-principles calculations and fluctuational electrodynamics, we quantitatively bridge the near-field radiative heat flux and anharmonic phonon scattering. While the inclusion of higher-order phonon scattering processes was known to result in more friction among phonons and reduce thermal conductivity, here we reveal, surprisingly, four-phonon scattering is favorable to NFRHT between polar dielectrics slabs and leads to 300-fold enhancement to the total heat flux for BAs and a 4-fold enhancement to that for BSb. By varying temperatures and gap distances, we show that smaller vacuum gap distance and lower temperature difference can make four-phonon scattering more dominant on the radiative heat transfer over three-phonon scattering. Meanwhile, considering the existence of optimum damping factor for NFRHT, we point out the possibility of four-phonon scattering to decrease NFRHT in certain other systems. Our work could open an avenue of phonon engineering and materials screening for near-field thermal modulation and management.

\begin{figure}[t]
\centering
\includegraphics[width=3.375in]{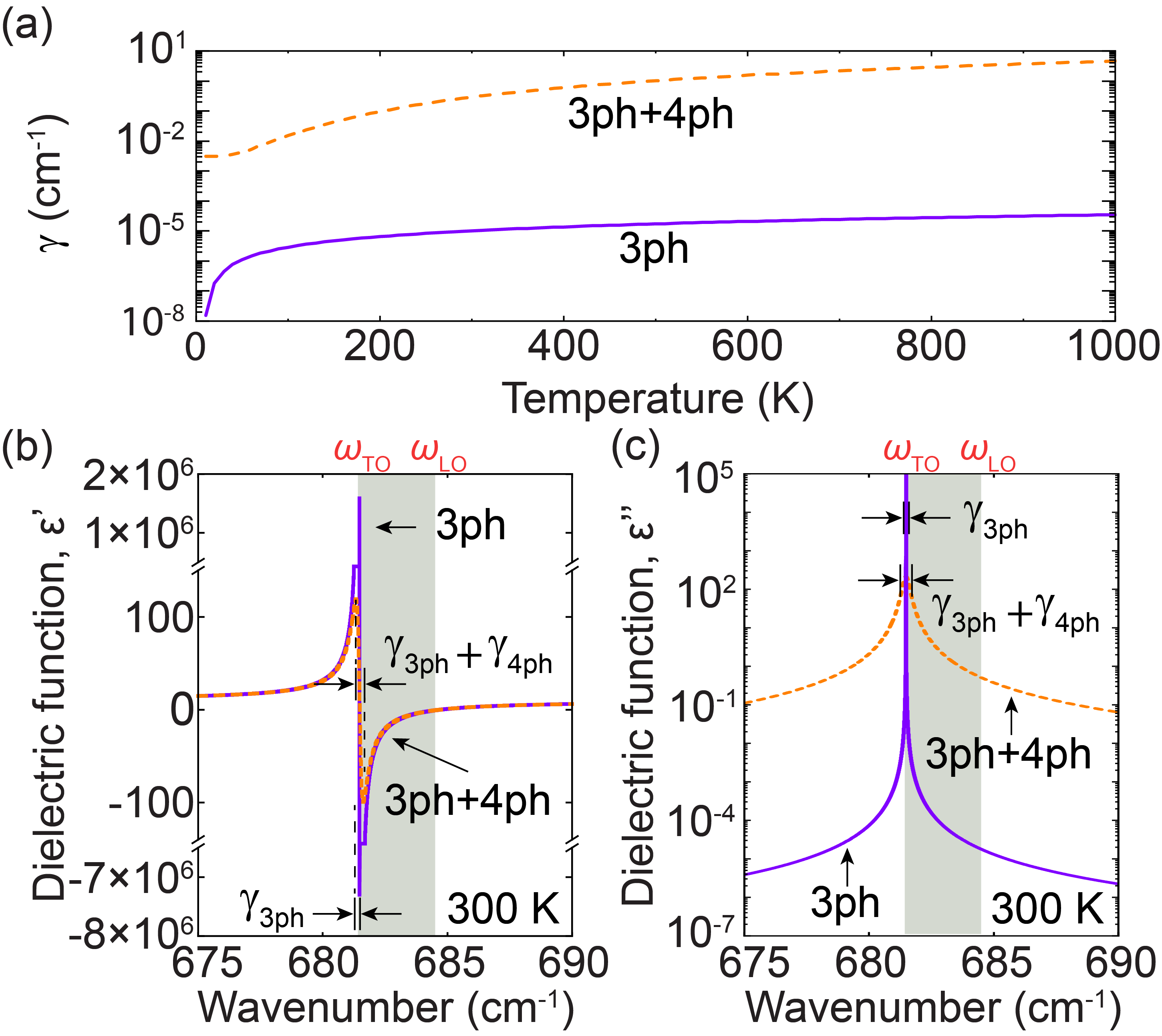}
\caption{\label{fig1} (a) The damping factors of isotopically pure BAs as a function of temperature with or without $\gamma_\mathrm{4ph}$~\cite{yang2020prb}. (b) The real and (c) imaginary parts of the dielectric function of isotopically pure BAs with or without $\gamma_\mathrm{4ph}$ at 300~K. The shaded area is the Reststrahlen band.}
\end{figure}

We start with the calculation of the temperature-dependent damping factor by summing the scattering rates of higher-order phonon scatterings:
\begin{equation}\label{eq:1}
\gamma(T)=\gamma_\mathrm{3ph}(T)+\gamma_\mathrm{4ph}(T)
\end{equation}
where $T$ is the material temperature at thermal equilibrium, and $\gamma$ represents the damping factor. The subscript 3ph and 4ph indicates the contributions from three-phonon and four-phonon scattering, respectively. Here we consider isotope pure BAs for simplicity, hence no phonon isotope scattering is needed. Figure~\ref{fig1}(a) shows the temperature-dependent damping factors of BAs before and after including four-phonon scattering. Although a monotonical increasing trend is shown as temperature increases with or without four-phonon scattering, the damping factor due to four-phonon scattering is 4$\sim$6 orders of magnitude larger than that due to three-phonon scattering. To satisfy the energy and momentum conservation in a phonon scattering process, three-phonon scattering, especially the process involving two acoustic phonons and one optical phonon (aao), is largely suppressed due to the large a-o gap so that $\gamma_\mathrm{3ph}$ is only $10^{-5}$~cm\textsuperscript{-1} at 300~K; while $\gamma_\mathrm{4ph}$ can reach 0.29~cm\textsuperscript{-1}. Therefore, four-phonon scattering dominates the damping force on the hypothetical optical oscillator used in the Lorentz oscillator model of the dielectric function~\cite{SupplMat}. The phonon frequencies and damping factors, obtained from first-principles calculations, are detailed in Supplementary Material~\cite{SupplMat}.

The dielectric functions of BAs are calculated with or without four-phonon scattering at 300 K, and the real ($\epsilon^{\prime}$) and imaginary parts ($\epsilon^{\prime\prime}$) of the dielectric functions around the Reststrahlen band are shown in Fig.~\ref{fig1}(b) and (c), respectively. We can clearly observe a strong oscillation of $\epsilon^{\prime}$ around the frequency, $\omega_\mathrm{r}$, when $\epsilon^{\prime}(\omega_\mathrm{r}) = 1$ in Fig.~\ref{fig1}(b)~\cite{zhang2020nano}. The peak value of $\epsilon^{\prime}$ without four-phonon scattering is 4-orders of magnitude larger than that with four-phonon scattering. The spectral width between the peak and valley values of $\epsilon^{\prime}$ is equivalent to the damping factor. As four-phonon scattering process dominates the anharmonic phonon scattering in BAs, this spectral width of $\epsilon^{\prime}$ of $\gamma_\mathrm{3ph}+\gamma_\mathrm{4ph}$ is much larger than that of $\gamma_\mathrm{3ph}$. As we can see from Fig.~\ref{fig1}(c), at the frequency far away from $\omega_\mathrm{r}$, $\epsilon^{\prime\prime}$ is negligible, which indicates that the absorption is only appreciable within an interval of damping factor around $\omega_\mathrm{r}$~\cite{zhang2020nano}. The full width at the half maximum (FWHM) of this absorption spectrum equals to the corresponding damping factor with or without four-phonon scattering included. Although $\epsilon^{\prime\prime}$ including four-phonon scattering has a wider FWHM, its peak value is much lower than that without including four-phonon scattering. This trade-off phenomenon implies, when the damping factor increases, more lattice vibration modes can potentially assist the light-matter interaction around $\omega_\mathrm{r}$, while the resonance strength is much suppressed. However, we observe a negative $\epsilon^{\prime}$ and a large $\epsilon^{\prime\prime}$ within the Reststralen band, where the normal reflectivity can reach close to unity due to the near-zero refractive index ($n$) and large extinction coefficient ($\kappa$)~\cite{yang2020prb}. Therefore, the weakly anharmonic polar materials are highly reflecting when interacting with propagating waves (far-field), while interacting with evanescent waves (near-field), especially surface waves, the materials become highly absorbing due to the phonon tunneling effect. 

\begin{figure}[t]
\centering
\includegraphics[width=3.375in]{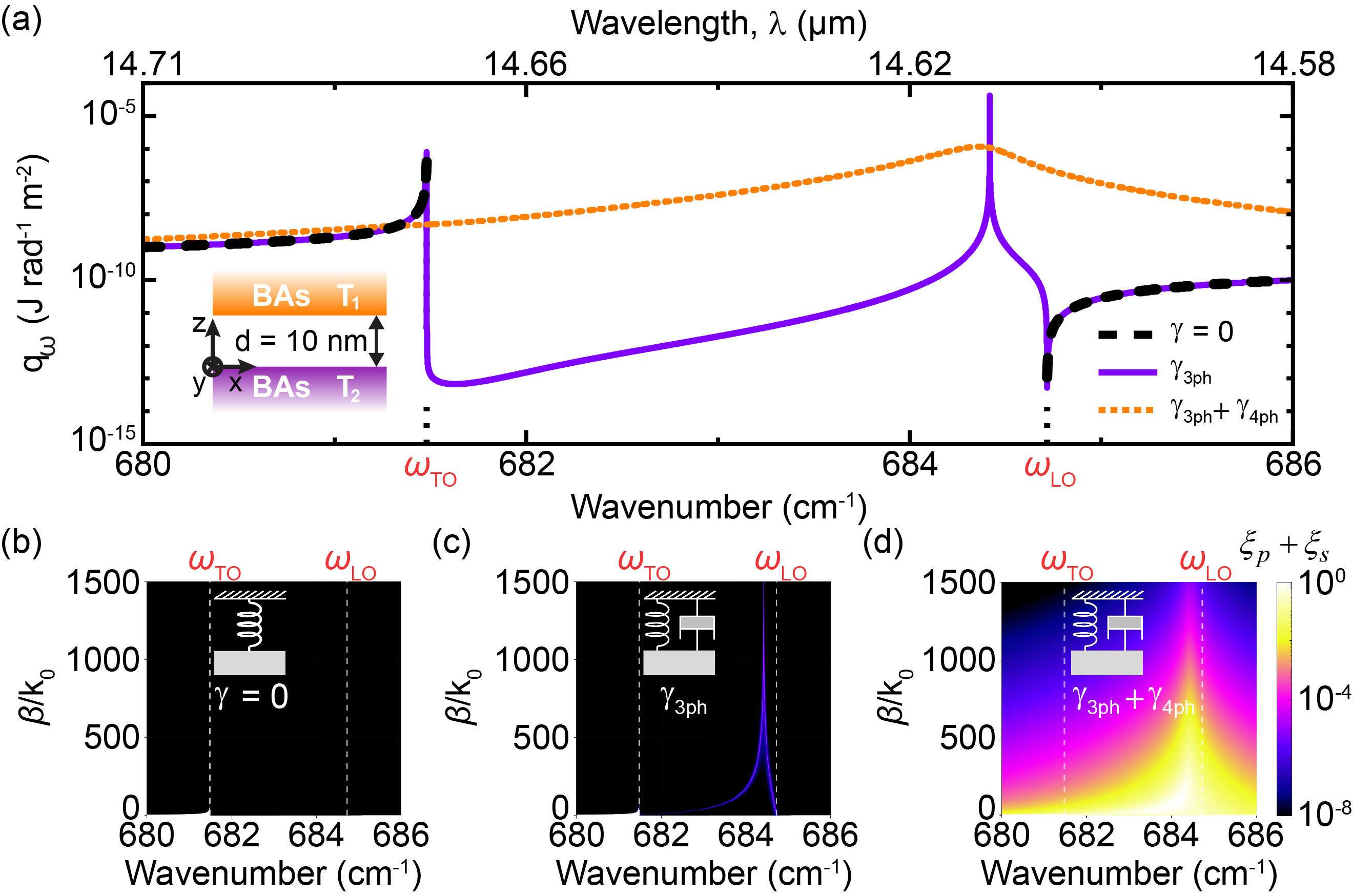}
\caption{\label{fig2} (a) The spectral heat fluxes between two BAs bulks with three different damping factors. The contour plots of the energy transmission coefficient ($\xi_s+\xi_p$) between two BAs bulks with (b) $\gamma=0$, (c) $\gamma_\mathrm{3ph}$, and (d) $\gamma_\mathrm{3ph}+\gamma_\mathrm{4ph}$. 
A mass-spring-damper system visualizes the Lorentz oscillations of the bonded crystals. With $\gamma = 0$, the mass-spring-damper system can be simplified as a mass-spring system. When $\gamma \neq 0$, the damper has a corresponding damping coefficient with or without $\gamma_\mathrm{4ph}$. The energy transmission coefficient is calculated with respect to the normalized parallel wavevector $\beta/k_0$ and angular frequency, with $k_0 = \omega/c$ ($\omega$ is the wavevector at vacuum, and $c$ is the speed of light). The insert plot in (a) indicates the configuration of the near-field radiation system with a 10 nm vacuum gap spacing and the temperatures of two BAs are set as $T_1 = 1000$~K and $T_2 = 300$~K, respectively. The vacuum gap distance is the default in this work unless stated otherwise.}
\end{figure}

We first present the spectral heat fluxes ($q_\omega$) with three different damping factors in Fig.~\ref{fig2}(a). A fluctuational electrodynamics formalism is applied to characterize NFRHT, as detailed in Supplemental Material~\cite{SupplMat}. A near-field radiation system composed of two bulk BAs with a hypothetical damping factor, $\gamma=0$, is calculated as the baseline, where the result indicates that the radiative heat transfer is completely prohibited within Reststrahlen band. Since $n = 0$, a perfect polar dielectric is perfect reflecting so that propagating waves are fully confined within BAs bulks. Additionally, the excitation of coupled SPhPs to transfer energy requires a nonzero damping factor based on the analysis of the local photon density of states~\cite{SupplMat}. In other words, no exchange of photons of any mode can occur between two perfect polar dielectric bulks, and phonon anharmonicity is necessary to facilitate the NFRHT mediated by SPhPs. To explicitly examine the four-phonon scattering effect on NFRHT, we compare $q_\omega$ with or without $\gamma_\mathrm{4ph}$. Within the Reststrahlen band, $q_\omega$ with $\gamma_\mathrm{3ph}$ is mostly 4$\sim$6 orders of magnitude smaller than that with $\gamma_\mathrm{3ph}+\gamma_\mathrm{4ph}$. The wider FWHM of $q_\omega$ with $\gamma_\mathrm{3ph}+\gamma_\mathrm{4ph}$ comparing to that with $\gamma_\mathrm{3ph}$ indicates that more optical phonon modes can participate in NFRHT through phonon scattering when four-phonon scattering is included. As $q_\omega \sim \mathrm{Im}(\epsilon_1)\mathrm{Im}(\epsilon_2)/|(\epsilon_1+1)(\epsilon_2+1)|^2$, the peak value of $q_\omega$ occurs at the resonant frequency ($\omega_\mathrm{res}$) when $\epsilon_1=-1$ or $\epsilon_2=-1$. $q_\omega$ at $\omega_\mathrm{TO}$ with $\gamma=0$ or $\gamma_\mathrm{3ph}$ is much higher than that with $\gamma_\mathrm{3ph}+\gamma_\mathrm{4ph}$, because $n(\omega_\mathrm{TO})$ with $\gamma=0$ or $\gamma_\mathrm{3ph}$ is much larger than that with $\gamma_\mathrm{3ph}+\gamma_\mathrm{4ph}$. Therefore, more frustrated modes can tunnel through the vacuum gap and enhance NFRHT. Same reason can be applied to explaining $q_\omega$ at $\omega_\mathrm{LO}$.

To investigate the contribution of each electromagnetic mode, we calculate the energy transmission coefficient ($\xi$) as a function of the normalized parallel wavevector and the angular frequency for different damping factors, as shown in Fig.~\ref{fig2}(b)-(d) for BAs. The calculation of $\xi$ and results of BSb can be found in Supplemental Material~\cite{SupplMat}. Each point on the contour plot represents a channel for radiative heat transfer at a specific parallel wavevector and angular frequency. The brighter color represents a higher photon transmission probability at this channel. As shown in Fig.~\ref{fig2}(b), we can clearly see that neither coupled evanescent wave nor propagating waves supports radiative heat transfer within the Reststrahlen band when $\gamma=0$. As indicated by the mass-spring system, no scattering acts as the friction force so that the surface resonance can never be excited through other photon modes at different frequencies. Moreover, as no local photon density of states supports the propagating waves~\cite{SupplMat}, the radiative heat transfer is completely forbidden within the Reststrahlen band. Outside the Reststrahlen band, heat can be transferred by propagating modes and frustrated modes, and no surface modes can be excited to support radiative heat transfer because the permittivity is not negative. As the damping factor only including three-phonon scattering is extremely small, the energy transmission coefficient is only significant along the dispersion relation of coupled SPhPs~\cite{SupplMat}. The peak value of $q_\omega$ with $\gamma_\mathrm{3ph}$ at $\omega_\mathrm{res}$ in Fig.~\ref{fig2}(a) is arisen from the surface modes excited at the parallel wavevector close to $1500k_0$. Comparing between Fig.~\ref{fig2}(c) and (d), the coupled SPhPs are widely and strongly excited when four-phonon scattering is considered into the damping factor. As the damping force of $\gamma_\mathrm{3ph}+\gamma_\mathrm{4ph}$ is much larger than that of $\gamma_\mathrm{3ph}$, the Lorentz oscillator with $\gamma_\mathrm{3ph}+\gamma_\mathrm{4ph}$ can interact with more modes at various frequencies so that more channels are activated to support the coupled SPhPs. Nonetheless, $q_\omega$ at $\omega_\mathrm{res}$ is dominated by the surface modes with large parallel wavevectors but the overall radiative heat transfer is enhanced when four-phonon scattering is involved. Therefore, NFRHT between two bulk BAs at nanoscale is enabled by phonon anharmonicity and greatly contributed by four-phonon scattering.

\begin{figure}[t]
\includegraphics[width=3.375in]{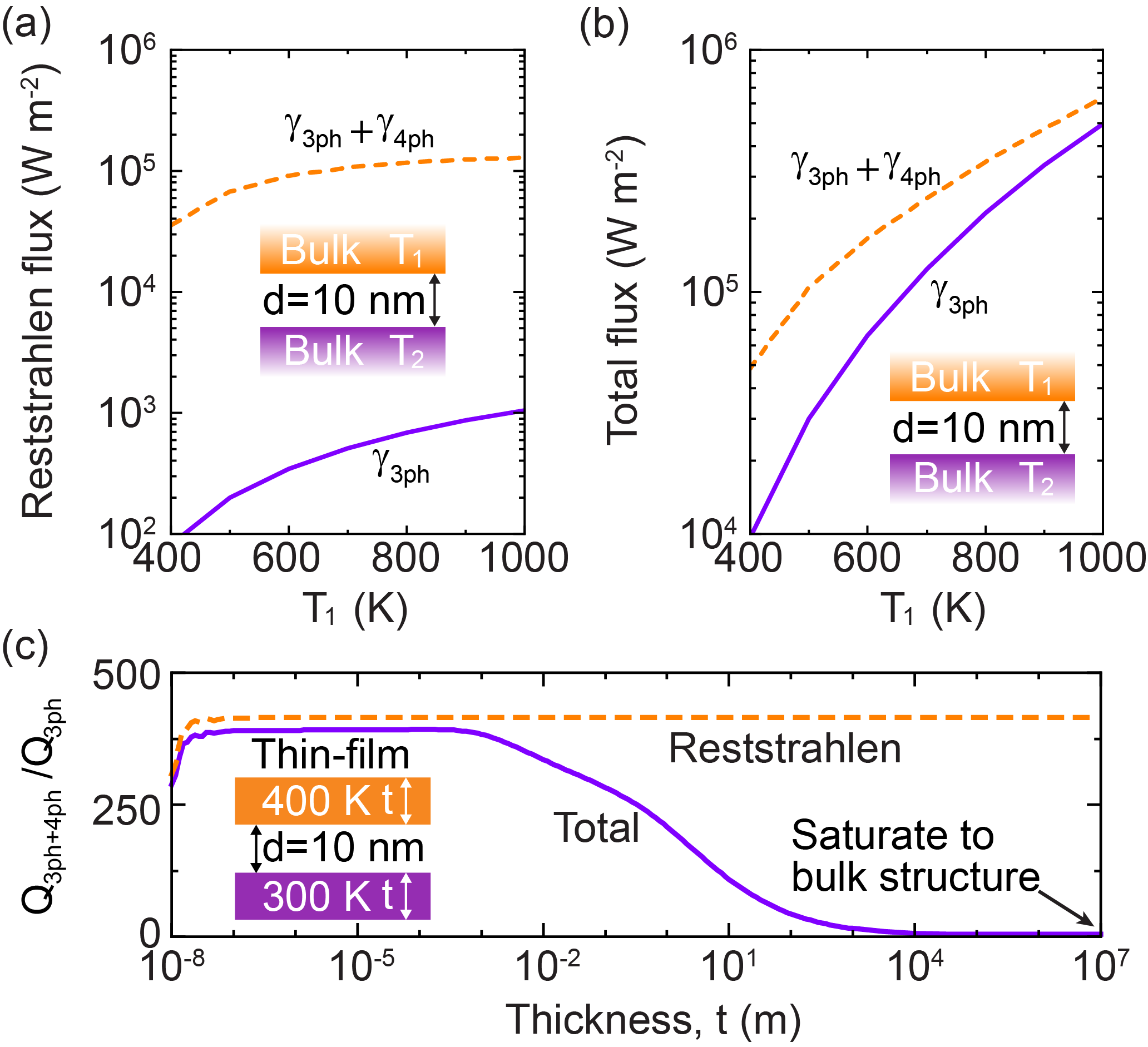}
\caption{\label{fig3} (a) The Reststrahlen band and (b) the total heat flux with or without considering four-phonon scattering with respect to $T_1$. Note that $T_2 = 300$~K. (c) The effect of four-phonon scattering on NFRHT with respect to the thicknesses of BAs films.}
\end{figure}

To examine the four-phonon scattering effect on NFRHT, we calculate the heat flux of the Reststrahlen band~($Q_\mathrm{Rest}$) and the total heat flux~($Q_\mathrm{t}$) for hot side temperatures ranging from 400~K to 1000~K. As shown in Fig.~\ref{fig3}(a), $Q_\mathrm{Rest}$ monotonically increases with respect to $T_1$. $Q_\mathrm{Rest}$ with four-phonon scattering is nearly 500 times higher than that with only three-phonon scattering when $T_1 = 400$~K, while this enhancement reduces to 100 times when $T_1 = 1000$~K. Although a higher temperature indicates a larger damping factor, more scattering processes do not necessarily enhance NFRHT linearly. The trade-off between the resonance peak value and the FWHM of the spectral heat flux implies that an optimal pair of damping factors may exist. The same trend of $Q_\mathrm{t}$ with respect to $T_1$ is also observed in Fig.~\ref{fig3}(b), while the enhancement brought by four-phonon scattering become much smaller, since the coupled SPhPs only contribute a fraction to the total radiative heat transfer. The enhancement due to four-phonon scattering decreases from 5~times to 1.3~times when $T_1$ increases from 400~K to 1000~K, driven by the closer agreement of $\gamma_\mathrm{3ph}+\gamma_\mathrm{4ph}$ at 400~K with the value at 300 K, facilitating better coupling of surface modes. Furthermore, the effect of four-phonon scattering becomes more apparent when reducing the thickness of the polar dielectric. To quantify this effect, we define a ratio $R_\mathrm{4ph}$ as $Q_\mathrm {3ph+4ph}/Q_\mathrm {3ph}$ for both total heat flux and Reststrahlen band flux. As shown Fig.~\ref{fig3}(c), the maximum $R_\mathrm{4ph}$ of total heat flux can surpass 300 since the radiative heat flux carried by propagating waves is suppressed in thin-film structures. Consequently, at small thicknesses, the ratio of the total heat flux closely matches that of the Reststrahlen band flux. However, as the thickness increases, the ratio of the total heat flux saturates and approaches the value observed in semi-infinite bulk structure (Fig.~\ref{fig3}(a) and (b)). Similar analysis and conclusions can be drawn for BSb~\cite{SupplMat}. Moreover, the significance of four-phonon scattering on NFRHT is investigated from near- to far-field regime in Supplementary Material~\cite{SupplMat}.

\begin{figure}[t]
\includegraphics[width=3.375in]{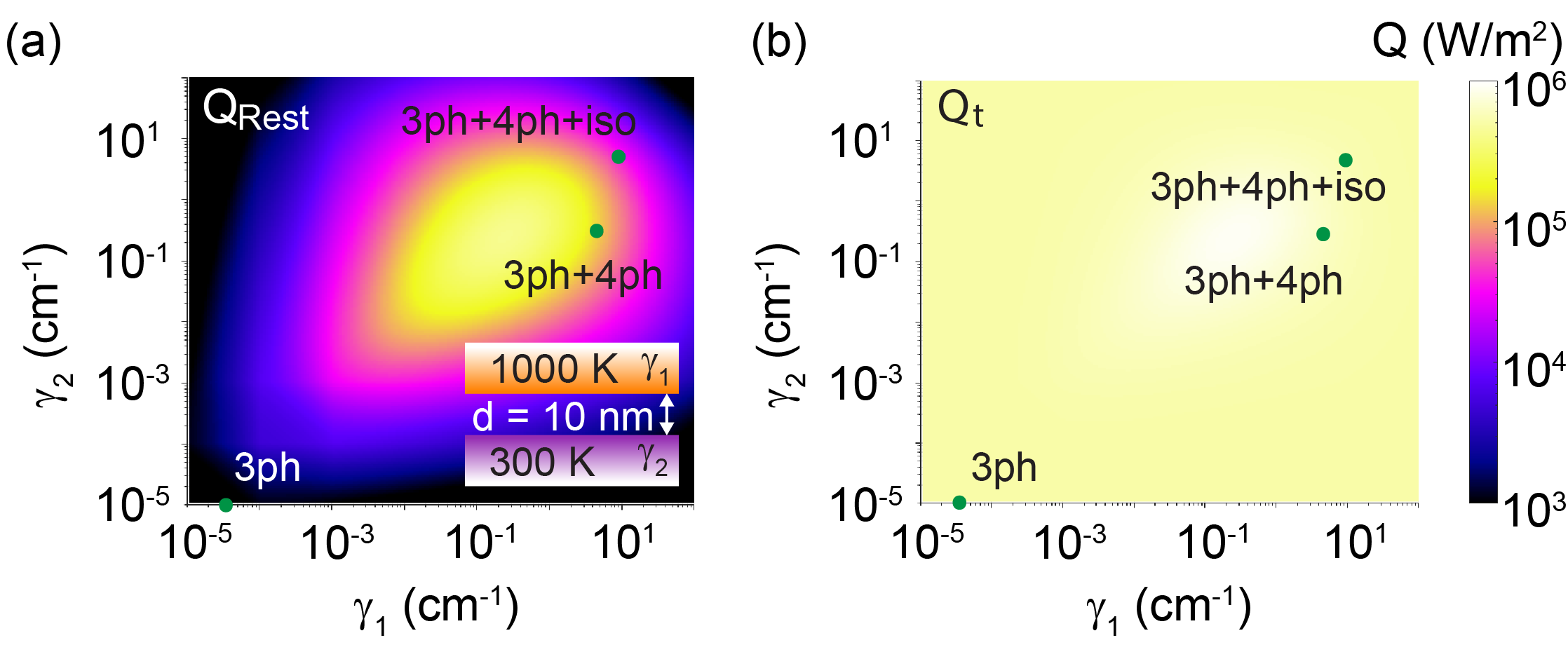}
\caption{\label{fig4}  The contour plot of the Reststrahlen band (a) and the total heat flux (b) with respect to the artificially tuned damping factors. Three specific points are marked to represent actual situations when three-phonon scattering, four-phonon scattering and phonon-isotope scattering are each added into the consideration.}
\end{figure}

We also perform a parametric sweep on the arbitrarily-tuned damping factors of two BAs bulks  at 1000 K and 300 K, and create two contour plots ~(Fig.~\ref{fig4}) of $Q_\mathrm{Rest}$ and $Q_\mathrm{t}$ to reveal the direct relation between phonon anharmonicity and radiative heat transfer. We mark three specific points representing the situations when considering 3ph, 3ph+4ph, and 3ph+4ph+iso, respectively. Clearly, there exist maximum values of both $Q_\mathrm{Rest}$ and $Q_\mathrm{t}$, and the corresponding damping factors of the hot and cold sides are both 0.21~cm\textsuperscript{-1}, which are at the order of 10\% of TO-LO splitting~\cite{Reddy2023}. When the damping factors are different for the hot and cold sides, the activated number of NFRHT channels is determined mostly by the smaller value of $\gamma_\mathrm{1}$ and $\gamma_\mathrm{2}$ due to the coupling effect of SPhPs, and the coupling strength of each channel is restricted by the larger value between the two damping factors. A pair of small damping factors indicate a narrow-band spectral heat flux but with a large peak at the resonant frequency. Therefore, as the pair of damping factors increase from $10^{-5}$~cm\textsuperscript{-1} to the optimal values, the peak spectral heat flux reduces but more SPhPs are coupled around $\omega_\mathrm{res}$ and $Q_\mathrm{Rest}$ increases. $Q_\mathrm{Rest}$ become more significant in radiative heat transfer, leading to a maximum in $Q_\mathrm{t}$. Once the pair of damping factors pass the optimal values, the resonant effect of coupled SPhPs becomes weaker and less radiative heat flux is carried by the surface modes. The three specific points representing the actual cases indicate that including four-phonon scattering can greatly enhance radiative heat transfer processes, while further including the isotope effect is likely to deteriorate NFRHT of BAs. However, isotope engineering has been shown to effectively tune radiative heat transfer~\cite{Bai2023PRB}. With recent estimate of fifth-order or even higher-order phonon scatterings~\cite{yang2022prb}, the residue damping factor results from these phonon scattering can potentially modify radiative heat transfer for both far- and near-field regime. The enhancement or deterioration of NFRTHT due to higher-order phonon scattering in different materials depends on whether $\gamma_\mathrm{4ph}$ can drive the damping factor towards the optimal value or away from it. Normally, four-phonon scattering can dominate NFRHT in materials where the three-phonon scattering is greatly suppressed. Conversely, if a material's three-phonon scattering rate is already near or at the optimum, adding four-phonon scattering may drive the total scattering rate pass the optimum hence reduce NFRHT. Nevertheless, the non-monotonic dependence of NFRHT on phonon scattering rates is distinctive from thermal conductivity where small phonon scattering rates are always required to increase the thermal conductivity for BAs and BSb.

In conclusion, we have quantitatively established that phonon anharmonicity is crucial for SPhPs-mediated radiative heat transfer. By considering four-phonon scattering alongside three-phonon scattering, we observe a remarkable enhancement of the total heat flux between BAs bulks by over 2 orders of magnitude, in contrast to its role in reducing thermal conductivity. The impact of four-phonon scattering on NFRHT becomes more pronounced with smaller vacuum gap distances and film thicknesses for both BAs and BSb. Moreover, through artificial tuning of the damping factor in BAs, we identify an optimal pair of damping factors that maximize radiative heat flux, indicating that the enhancement of NFRHT through four-phonon scattering requires materials with a suppressed damping factor due to three-phonon scattering. Therefore, we also suggest that four-phonon scattering to decrease NFRHT in certain other systems. Our work not only unveils the relationship between phonon scattering and NFRHT in weakly anharmonic polar dielectrics, highlighting the potential of phonon engineering to control radiative heat transfer at nanoscales. This study opens a promising avenue for near-field thermal regulation and management techniques.
\begin{acknowledgments}
D.F. and X.R. acknowledge partial support from the US National Science Foundation through award 2102645. Z.H. and X.R. acknowledge partial support from the US National Science Foundation through award 2015946.
\end{acknowledgments}

\bibliography{apssamp}

\end{document}


\title{Supplemental Material for ``Four Phonon-Dominated Near-Field Radiation in Weakly Anharmonic Polar Materials"}

\author{Dudong Feng}
\affiliation{%
School of Mechanical Engineering and the Birck Nanotechnology Center, Purdue University, West Lafayette, Indiana 47907-2088, USA
}%


\author{Xiaolong Yang}
\altaffiliation[Current at ]{College of Physics, and Center of Quantum Materials and Devices, Chongqing University, Chongqing 401331, China}

\affiliation{%
 School of Mechanical Engineering and the Birck Nanotechnology Center, Purdue University, West Lafayette, Indiana 47907-2088, USA
}%

\author{Zherui Han}
\affiliation{%
 School of Mechanical Engineering and the Birck Nanotechnology Center, Purdue University, West Lafayette, Indiana 47907-2088, USA
}%

\author{Xiulin Ruan}%
 \email{ruan@purdue.edu}
\affiliation{%
School of Mechanical Engineering and the Birck Nanotechnology Center, Purdue University, West Lafayette, Indiana 47907-2088, USA
}%


\date{\today}

\maketitle
\pagebreak

\section{The effect of four-phonon scattering on NFRHT between Boron Antimonide}

\begin{figure}[h]
\centering
\includegraphics[width=3.375in]{fig s1.png}
\caption{\label{figs1} (a) The damping factors of isotopically pure BSb as a function of temperature with or without $\gamma_\mathrm{4ph}$. (b) The real and (c) imaginary parts of the dielectric function of isotopically pure BSb with or without $\gamma_\mathrm{4ph}$ at 300~K. The shaded area is the Reststrahlen band.}
\end{figure}

\begin{figure*}[h]
\centering
\includegraphics[width=\textwidth]{fig s2.png}
\caption{\label{figs2} (a) The spectral heat fluxes between two BSb bulks with three different damping factors. The contour plots of the energy transmission coefficient ($\xi_s+\xi_p$) between two BSb bulks with (b) $\gamma_\mathrm{3ph}$, and (c) $\gamma_\mathrm{3ph}+\gamma_\mathrm{4ph}$. The insert plot in (a) indicates the configuration of the near-field radiation system. The vacuum gap spacing is set to be 10 nm and the temperatures of two BSb are set as $T_1 = 1000$~K and $T_2 = 300$~K, respectively. }
\end{figure*}

\begin{figure}[h]
\centering
\includegraphics[width=3.375in]{fig s3.png}
\caption{\label{figs3} (a) The Reststrahlen band and (b) the total heat fluxwith or without considering four-phonon scattering with respect to $T_1$.(c) The effect of four-phonon scattering on NFRHT with respect to the thicknesses of BSb films.}
\end{figure}
 \clearpage

\section{Computational details of the phonon linewidth}
The calculation of the phonon linewidth and frequencies requires both harmonic and anharmonic interatomic force constants (IFCs), which are obtained from density functional calculations within the localized density approximation for BAs and the Perdew–Burke–Ernzerhof generalized gradient approximation for BSb \cite{KRESSE1996CMS, Hafner1993prb, Perdew1996prl}, as implemented in the Vienna Ab initio Simulation Package (VASP)~\cite{KRESSE1996CMS}. The kinetic energy cutoff of plane-wave basis is set as 520~eV, and the Brillouin zone (BZ) are sampled using the Monkhorst-Pack grid of $20\times20\times20$ for the primitive unit cell. The lattices and internal atomic positions are fully relaxed with the total energy convergence threshold of $10^\mathrm{-10}$~eV and a criterion of $10^\mathrm{-6}$~eV/\AA for the forces on each atom, respectively. The harmonic IFCs are calculated using $5\times5\times5$ supercells and a $4\times4\times4$ Monkhorst-Pack grid via the finite displacement method, by employing the Phonopy~\cite{Tanaka2008prb}. The non-analytical corrections are applied to the dynamical matrix to take into account long-range electrostatic interactions. The third-order IFCs are calculated through Thirdorder~\cite{Shengbte} considering up to the fifth-nearest neighbor, with $4\times4\times4$ supercells and a $4\times4\times4$ Monkhorst-Pack grid. The fourth-order IFCs are calculated by using Fourthorder~\cite{han2022CPC}, in which the second-nearest neighbors are considered using $4\times4\times4$ supercells. The phonon linewidth is calculated with integrations using $16\times16\times16$ and $20\times20\times20$ q-mesh in the BZ for BAs and BSb, respectively, which affords the converged phonon scattering rates~\cite{feng2017prb}. 

\section{Lorentz model and near-field radiation between polar dielectrics}
The dielectric function of polar dielectrics can be described by the Lorentz model~\cite{zhang2020nano}.
\begin{equation}\label{eq:2}
\epsilon(\omega)=\epsilon_{\infty} \left( 1+\frac{\omega_{\mathrm{LO}}^{2}-\omega_{\mathrm{TO}}^{2}}{\omega_{\mathrm{TO}}^{2}-\omega^{2}-i\omega\gamma} \right)
\end{equation}
For BAs, $\epsilon_\infty = 9.83$, representing the dielectric constant at the high-frequency limit, and $\omega$ is the photon frequency. $\omega_\mathrm{TO} = 681.5$~cm\textsuperscript{-1} and $\omega_\mathrm{LO} = 684.7$~cm\textsuperscript{-1}. For BSb, $\omega_\mathrm{TO} = 587.9$~cm\textsuperscript{-1}, $\omega_\mathrm{LO} = 607.3$~cm\textsuperscript{-1}, and $\epsilon_\infty = 10.94$.

A fluctuational electrodynamics formalism is applied to characterizing the near-field radiative heat transfer between two polar dielecrics. The heat flux of the Reststrahlen band and the total heat flux can be expressed as following, respectively.
\begin{equation}\label{eq:S1}
Q_\mathrm{Rest} = \int_{\omega_\mathrm{TO}}^{\omega_\mathrm{LO}}q_{\omega}d\omega
\end{equation}
and
\begin{equation}\label{eq:S2}
Q_\mathrm{t} = \int_{0}^{\infty}q_{\omega}d\omega
\end{equation}
The spectral heat flux per unit area is calculated by~\cite{zhang2020nano}
\begin{equation}\label{eq:S3}
q_\omega = \frac{[\Theta(\omega,T_1)-\Theta(\omega,T_2)]}{4\pi^2}\int_{0}^{\infty}\sum_{m=s,p}\xi_\mathrm{m}(\omega,\beta)\beta{d\beta}
\end{equation}
where $\Theta(\omega,T)=\hbar\omega[\exp(\hbar\omega/k_\mathrm{B}T)-1]^{-1}$ is the mean energy of a Planck oscillator at an angular frequency $\omega$ with $\hbar$ and $k$\textsubscript{B} being the reduced Planck constant and Boltzmann constant, respectively. $T_1$ and $T_2$ represents the hot and cold side temperatures, respectively. $\beta=\sqrt{k_x^2+k_y^2}$ is the magnitude of the parallel wavevector in the x-y plane. $\xi_\mathrm{m}(\omega,\beta)$ is the energy transmission coefficient, and subscript m represents either the transverse electric waves (\emph{s}-polarization) or transverse magnetic waves (\emph{p}-polarization). Depends on the geometric structures, the calculation of energy $\xi_\mathrm{m}(\omega,\beta)$ can be found for both semi-infinite bulk structure and thin-film structure~\cite{zhang2020nano}. A cut-off parallel wavevector is required to numerically calculate the integral of Eq.~\ref{eq:S3}. According to Ref.~\cite{basu2009jap}, $\pi/d_c$ is usually a safe value for the spectral heat flux to converge, where $d_c$ is the lattice distance. However, the maximum value of the parallel wavevector of the excited surface modes can easily exceed $1000k_0$ as shown in Fig.~2 for a polar dielectric with an extremely small damping factor. Most of the nonzero energy transmission coefficients locate along the dispersion curves of the coupled SPhPs. Therefore, an adaptive cutoff of the parallel wavevector is chosen to achieve the convergence with a reasonable computational cost. The convergence analysis is performed for different damping factors. A finer mesh is required in the frequency and parallel wavevector space when the material possess a smaller damping factor (~$10^{-5}$~cm\textsuperscript{-1}). A $10^\mathrm{-6}\times10^\mathrm{-6}$ $\beta-\omega$ mesh is used to calculate the spectral heat flux in the Reststrahlen band for $\gamma_\mathrm{3ph}$.

The dispersion relation of the coupled SPhPs in the vacuum sandwiched between two identical BAs bulks can be expressed by~\cite{zhang2020nano}

\begin{equation}\label{eq:S4}
\tanh{ik_{z0}d}\left(\frac{k_{z0}^{2}}{\epsilon_{0}^{2}}+\frac{k_{z1}k_{z2}}{\epsilon_1\epsilon_2}\right)= \frac{k_{z0}}{\epsilon_{0}}\left(\frac{k_{z1}}{\epsilon_1}+\frac{k_{z2}}{\epsilon_2}\right)
\end{equation}
where $k_{z0}=\sqrt{k_0^2-\beta^2}$, $k_{z1}=\sqrt{\epsilon_1k_0^2-\beta^2}$ and $k_{z2}=\sqrt{\epsilon_2k_0^2-\beta^2}$.

\begin{figure}[t]
\includegraphics[scale=1]{fig s4.png}
\caption{\label{figs4} The dispersion relation of the coupled SPhPs between two BAs bulks with or without four-phonon scattering. Note that temperature-dependent damping factor is considered.}
\end{figure}

To better describe the effect of four-phonon scattering on the pattern of the energy transmission coefficient, we calculate the dispersion relation of NFRHT with or without $\gamma_\mathrm{4ph}$. Compare Fig. 2(c) with Supplementary Fig.~\ref{figs4}, we can see that the active channels of radiative heat transfer on the contour plot distribute along the dispersion relation of coupled SPhPs with $\gamma_\mathrm{3ph}$. Since $\gamma_\mathrm{3ph}$ is on the order of $10^{-5}$~cm\textsuperscript{-1} from 300~K to 1000~K, the coupled SPhPs can be excited at a really large parallel wavevector close to $1500k_0$. As $\gamma_\mathrm{3ph}$ below 1000~K is very small, the dispersion relation can be divided into two modes, a low frequency symmetric mode and a high frequency antisymmetric mode~\cite{zhang2020nano}. At the low frequency symmetric mode, the surface charges are symmetric and magnetic fields at the interface are in phase. At the high antisymmetric mode, the surface charges are asymmectric and the magnetic field are out of phase at the interface. However, for the case of $\gamma_\mathrm{3ph}+\gamma_\mathrm{4ph}$, the two modes are hardly distinguishable because the damping factor is too large so that the peak is not as sharp as the case of $\gamma_\mathrm{3ph}$. The dispersion curve of coupled SPhPs with $\gamma_\mathrm{3ph}+\gamma_\mathrm{4ph}$ also exhibits the distribution of the major radiative transfer modes with high energy transmission coefficients.

\section{The excitation of SPhPs in vacuum}

The local photon density of states above the surface of a polar dielectric bulk explicitly indicates the existence of different electromagnetic modes. At a specific location $z$, the local photon density of states can be expressed as:
\begin{equation}\label{eq:S5}
D(z,\omega) = D_\mathrm{prop}(\omega)+D_\mathrm{evan}(z,\omega)
\end{equation}
where
\begin{equation}\label{eq:S6}
D_\mathrm{prop}(\omega) = \int_{0}^{k_0}\frac{\omega}{2\pi^2c_0k_{z0}}\left(2-\rho_\mathrm{0m}^{s}-\rho_\mathrm{0m}^{p}\right)d\beta
\end{equation}
and
\begin{equation}\label{eq:S7}
D_\mathrm{evan}(\omega) = \int_{k_0}^{\infty}\frac{\exp{(2ik_{z0}z)}}{-2i\pi^2\omega k_{z0}}\left[\mathrm{Im}(r_\mathrm{0m}^{s})+\mathrm{Im}(r_\mathrm{0m}^{p})\right]\beta^{3}d\beta
\end{equation}
Here, $\rho_{0m} = \lvert{r_\mathrm{0m}}\rvert^2$, where $r_\mathrm{0m}$ is the Fresnel reflection coefficient. m can be media 1 and 2.

\begin{figure}[t]
\includegraphics[scale=0.85]{fig s5.png}
\caption{\label{figs5} (a) The location of the calculated local photon density of states. The local photon density of states with (b) $\gamma = 0$, (c) $\gamma_\mathrm{3ph}$ at 300~K, (d) $\gamma_\mathrm{3ph}$ at 1000~K, (e) $\gamma_\mathrm{3ph}+\gamma_\mathrm{4ph}$ at 300 K, (f) and $\gamma_\mathrm{3ph}+\gamma_\mathrm{4ph}$ at 1000~K.  The contribution of propagating modes and evanescent modes are distinguished by purple and orange color, respectively.}
\end{figure}

 As shown in Supplementary Fig.~\ref{figs5}(a), the local photon density of states is calculated at 5~nm away from the surface of BAs bulk. Within the Reststrahlen band, no local photon density of states exists, which indicates that neither propagating modes nor evanescent modes are allowed to support radiative heat transfer when $\gamma = 0$. Outside the Reststrahlen band, both propagating and evanescent modes are generated to carry heat. When only considering three-phonon scattering, the local photon density of states is mostly supported by evanescent modes within the Reststrahlen band, since the propagating modes are confined with near-zero refractive index. The peaks around $\omega_\mathrm{TO}$ in Supplementary Fig.~\ref{figs5}(c) and (d) result from an extremely large refractive index. The peaks inside the Reststrahlen band in Supplementary Fig.~\ref{figs5}(c) and (d) arise from the excited SPhPs, since the refractive index is near-zero and all propagating waves generated inside the BAs can never be totally confined. If we include four-phonon scattering, the local photon density of states can be contributed by propagating modes since the refractive index is no longer close to zero. If comparing Supplementary Fig.~\ref{figs5}(c) with (d), or Supplementary Fig.~\ref{figs5}(d) with (f), we can see that the local photon density of states without four-phonon scattering is larger than that with four-phonon scattering at two frequencies ($\omega_\mathrm{TO}$ and $\omega_\mathrm{res}$). However, the local photon density of states with four-phonon scattering is 3 orders of magnitude larger than that without four-phonon scattering at most frequencies. Overall, if we compare propagating modes with evanescent modes for Supplementary Fig.~\ref{figs5}(b)-(f), the local photon density of states supported by propagating modes are negligible. Therefore, we can claim that most of thermal energy is transferred by near-field radiation at nanoscale gap distance~($<$~10~nm). Compared to three-phonon scattering, four-phonon scattering significantly support the local photon density of states close to the BAs surface.

\section{The effect of the vacuum gap distance on near-field radiative heat transfer}
\begin{figure}
\includegraphics[scale=0.98]{fig s6.png}
\caption{\label{figs6} (a) The Reststrahlen band and (b) the total heat flux in BAs with or without considering four-phonon scattering with respect to the vacuum gap distance. (c) The Reststrahlen band and (d) the total heat flux in BSb with or without considering four-phonon scattering with respect to the vacuum gap distance. The temperatures of hot and code sides are set as $T_1 = 400$~K and $T_2 = 300$~K, respectively.}
\end{figure}
We also vary the vacuum gap distance to see the role of four-phonon scattering on NFRHT at different working regimes. As the vacuum gap distance increases, the near-field enhancement on the radiative heat transfer is drastically reduced for the heat flux of the Reststrahlen band and full-spectrum for both BAs and BSb. As shown in Supplementary Fig.~\ref{figs6}(a) and (c), the heat fluxes of the Reststrahlen band for $\gamma_\mathrm{3ph}$ and $\gamma_\mathrm{3ph}+\gamma_\mathrm{4ph}$ decrease as the vacuum gap distance increases. As the vacuum gap distance reaches 0.1~mm (far-field regime), the coupled SPhPs carry nearly zero heat flux ($1\times10^{-5}~\mathrm{W/m^{2}}$) when only three-phonon scattering is considered, since the photons are highly confined at small damping factor. The heat flux of the Reststrahlen band reduces to a saturated value, which represents the radiative heat flux at far-field limit. However, if we examine the total heat flux, the four-phonon scattering effect is only significant when the vacuum gap distance is less than 100~nm for BAs and 200~nm for BSb. As shown in Supplementary Fig.~\ref{figs6}(b), the total heat flux of $\gamma_\mathrm{3ph}$ and $\gamma_\mathrm{3ph}+\gamma_\mathrm{4ph}$ become overlapping when the vacuum gap increases. Because the SPhPs can only dominate NFRHT at extremely small vacuum gap distance. However, this is the case for the hot side temperature at 400~K. The impact of four-phonon scattering can potentially be amplified when the hot side temperature is even lower.

\bibliography{apssamp}